\documentclass[twocolumn]{aastex62}

\submitjournal{\apj}

\begin{document}

\title{Revisiting the Li abundances of Stars with and without Detected Planets from the High Resolution Spectroscopy}

\author[0009-0000-2934-4067]{Jinxiao Qin}
\affiliation{Department of Physics, Hebei Normal University, Shijiazhuang 050024, People's Republic of China}
\affiliation{Institute for Frontiers in Astronomy and Astrophysics, Beijing Normal University, Beijing 102206, China}
\affiliation{Guo Shoujing Institute for Astronomy, Hebei Normal University, Shijiazhuang 050024, People's Republic of China}

\author[0000-0002-8609-3599]{Hong-Liang Yan}
\affiliation{CAS Key Laboratory of Optical Astronomy, National Astronomical Observatories, Beijing 100101, China}
\affiliation{Institute for Frontiers in Astronomy and Astrophysics, Beijing Normal University, Beijing 102206, China}
\affiliation{School of Astronomy and Space Science, University of Chinese Academy of Sciences, Beijing 100049, People's Republic of China}

\author[0000-0003-1359-9908]{Wenyuan Cui} 
\affiliation{Department of Physics, Hebei Normal University, Shijiazhuang 050024, People's Republic of China}
\affiliation{Guo Shoujing Institute for Astronomy, Hebei Normal University, Shijiazhuang 050024, People's Republic of China}

\author[0000-0002-0349-7839]{Jian-Rong Shi}
\affiliation{CAS Key Laboratory of Optical Astronomy, National Astronomical Observatories, Beijing 100101, China}
\affiliation{School of Astronomy and Space Science, University of Chinese Academy of Sciences, Beijing 100049, People's Republic of China}

\author[0000-0002-1027-0990]{Subo Dong}
\affiliation{Department of Astronomy, School of Physics, Peking University, Yiheyuan Rd. 5, Haidian District, Beijing, China, 100871}
\affiliation{Kavli Institute of Astronomy and Astrophysics, Peking University, Yiheyuan Rd. 5, Haidian District, Beijing, China, 100871}

\author[0000-0001-5193-1727]{Shuai Liu}
\affiliation{CAS Key Laboratory of Optical Astronomy, National Astronomical Observatories, Beijing 100101, China}
\affiliation{School of Astronomy and Space Science, University of Chinese Academy of Sciences, Beijing 100049, People's Republic of China}
\affiliation{Instituto de Astrof\'{i}sica de Canarias, V\'{i}a L\'{a}ctea, 38205 La Laguna, Tenerife, Spain}

\author[0000-0002-1619-1660]{Zeming Zhou}
\affiliation{Institute for Frontiers in Astronomy and Astrophysics, Beijing Normal University, Beijing 102206, China}
\affiliation{Department of Astronomy, Beijing Normal University, Beijing 100875, China}

\author[0000-0002-8888-6648]{Miao Tian}
\affiliation{Department of Physics, Hebei Normal University, Shijiazhuang 050024, People's Republic of China}
\affiliation{Guo Shoujing Institute for Astronomy, Hebei Normal University, Shijiazhuang 050024, People's Republic of China}

\author[0009-0008-8978-7052]{Zhenyan Huo}
\affiliation{Department of Physics, Hebei Normal University, Shijiazhuang 050024, People's Republic of China}
\affiliation{Guo Shoujing Institute for Astronomy, Hebei Normal University, Shijiazhuang 050024, People's Republic of China}

\author[0000-0003-3240-1688]{Xiangsong Fang}
\affiliation{CAS Key Laboratory of Optical Astronomy, National Astronomical Observatories, Beijing 100101, China}

\author[0000-0002-2510-6931]{Jinghua Zhang}
\affiliation{South-Western Institute for Astronomy Research, Yunnan University, Chenggong District, Kunming 650500, China}

\author[0000-0002-6647-3957]{Chunqian Li}
\affiliation{CAS Key Laboratory of Optical Astronomy, National Astronomical Observatories, Beijing 100101, China}
\affiliation{School of Astronomy and Space Science, University of Chinese Academy of Sciences, Beijing 100049, People's Republic of China}

\author[0000-0001-6898-7620]{Mingyi Ding}
\affiliation{CAS Key Laboratory of Optical Astronomy, National Astronomical Observatories, Beijing 100101, China}
\affiliation{School of Astronomy and Space Science, University of Chinese Academy of Sciences, Beijing 100049, People's Republic of China}

\author[0000-0003-3116-5038]{Song Wang}
\affiliation{Key Laboratory of Optical Astronomy, National Astronomical Observatories, Chinese Academy of Sciences, Beijing 100101, China}
\affiliation{Institute for Frontiers in Astronomy and Astrophysics, Beijing Normal University, Beijing 102206, China
}

\author[0000-0003-3474-5118]{Henggeng Han}
\affiliation{Key Laboratory of Optical Astronomy, National Astronomical Observatories, Chinese Academy of Sciences, Beijing 100101, China}

\correspondingauthor{Hong-Liang Yan and Wenyuan Cui}
\email{hlyan@nao.cas.cn, cuiwenyuan@hebtu.edu.cn}

\begin{abstract}

Whether the presence of planets affects the lithium (Li) abundance of their host stars is still an open question. To investigate the difference of the Li abundance between planet-host stars (HS) and isolated stars (IS) with no detected planets, we analyze a large sample of stars with temperatures ranging from $4600$ to $6600$\,K and metallicity ranging from $-0.55$ to $+0.50$. The sample consists of $279$ HS whose spectra were taken from the California-{\it Kepler} Survey (CKS), which followed up planets detected by {\it Kepler}, and 171 IS whose spectra were taken from the Keck archive. The non-local thermodynamic equilibrium (non-LTE) effects were taken into consideration. It is found that the distribution of Li abundances in both the HS and IS groups are generally consistent with each other. This suggests that the presence of {\it Kepler}-like planets does not have a significant impact on Li depletion. We also found that the non-LTE corrections can not be neglected for stars with A(Li) over $\sim 2.5$\,dex. 

\end{abstract}

\keywords{Stellar abundances$-$Planet hosting stars$-$Stellar spectral lines}
\section{Introduction}
Lithium is a commonly-used tracer in stellar physics. It is so fragile that it can be destroyed in stellar interiors at temperatures of $2.5-3\times 10^{6}$ \citep{1990ApJS...74..501P,1997ApJ...491..339S,2021FrASS...8....6R}, causing Li depletion in stars.
It has been suspected that Li depletion in planet-host stars (HS) can be affected by planetary formation and evolutionary processes.
The interactions with continued planetary disks may slow down the rotation of the HS. This process results in a high rotational difference between the core and the convective envelope, enhancing the mixing between the two components, which causing more severe Li depletion in the slower-rotating stars \citep{2008A&A...489L..53B,2002A&A...386.1039M}.
Therefore, Li abundance can provide information on the angular-momentum evolution of the stellar system.
Due to these properties, Li plays a unique role in studying HS.

Since the discovery of the first exoplanet, numerous studies have been conducted to investigate the potential connection between Li abundances and planetary systems\citep{1994Sci...264..538W, 2016MNRAS.462.1563M, 2012A&A...547A..36A, 2014A&A...564L..15A}. \citet{1997AJ....113.1871K} firstly proposed that Li depletion of HS is associated with their planets. Later, \cite{2004A&A...414..601I} found that HS are significantly more Li-poor than IS within the effective temperature range from $5600$ to $5850$\,K. 
However, for higher effective temperatures ($5850-6350$\,K), there is no notable distinction in the distribution of Li abundance between HS and IS. \citet{2006AJ....131.1816C} analyzed Li abundances of 16 HS with the effective temperature range of $5600-5900$\,K in the main–sequence stage, and they compared them with $20$ normal stars of the same range of effective temperature and metallicity.
Their result is consistent with \cite{2004A&A...414..601I} that there is a higher probability for HS to have depleted Li abundances in $5600-5900$\,K. 
Later, \citet[][hereafter I09]{2009Natur.462..189I} reported that the solar-type HS show enhanced Li depletion compared with IS of the same type and age, especially for the stars with effective temperature range between $5700-5850$\,K. This study investigated $70$ confirmed HS and $381$ IS.

However, other studies show that the Li depletion in HS is the same as that in the IS. 
\cite{2000MNRAS.316L..35R} showed that the Li abundances of HS are not significantly different from those of IS for the same age and stellar parameters. 
They compared the Li abundances of HS with the known members in open clusters and field IS. 
After excluding young stars and sub-giants, they found that there was no robust evidence that HS have lower Li abundances than IS of similar parameters.
\cite{2010A&A...519A..87B} studied a sample of $117$ solar-like stars, and obtained their atmospheric parameters and Li abundances.
\cite{2010A&A...519A..87B} used the same HS sample given by I09, but a different IS sample which was selected within a $2\sigma$ range in effective temperatures ($T_{\rm{eff}}$), surface gravities ($\log{g}$), and metallicities ([Fe/H]) around HS parameters space for comparison. 
They found that the HS did not show more Li depletion than the IS did and stress that Li depletion is dominated by age \citep[see][Fig.5]{2010A&A...519A..87B}. 
\citet[][hereafter L24]{2024A&A...684A..28L} analyzed the Li abundances of 1332 FGK main-sequence stars, among which 257 stars have detected planets. They considered influence of tides, orbital decay and other physical properties of the host stars, and found no evidence that planetary systems affect Li depletion or enrichment. In addition, many other works have also found the similar conclusion\citep[e.g.][]{2006AJ....131.3069L,2010ApJ...724..154G,2019MNRAS.487.3162C}.

Launched in $2009$, NASA's {\it Kepler} satellite dramatically increased the number of exoplanet detections by many thousands with the transit method, giving us the opportunity to revisit the long-standing question on the Li abundance with planetary hosts.
On one hand, more and more spectroscopic observations to the HS can be conducted by ground-based telescopes.
For example, the California-{\it Kepler} Sky Survey (CKS) is a massive spectroscopic survey conducted with the Keck/HIRES to the Kepler Objects of Interest (KOIs) \citep{2017AJ....154..107P,2017AJ....154..109F,2017AJ....154..108J}. The CKS has provided a vast collection of high-resolution HS spectra, facilitating the expansion of Li abundance determination in HS from several tens to hundreds. 
Moreover, the radial velocity monitoring helps to purify the IS sample as well. Some of the IS used in previous studies were found to host planets by subsequent observations.

This paper aims to investigates the relationship between stellar Li abundance and presence of planets in a large sample of HS and IS. We present the non-LTE Li abundances for $450$ FGK stars in total, of which $279$ stars are confirmed HS by the {\it Kepler} mission, and $171$ stars are currently IS by the \citet{2016ApJS..225...32B} and \citet{2020ApJ...896...64L}. All the spectra of our sample were taken by the same telescope and instrument (Keck/HIRES) under similar resolutions.
The outline of this paper is as follows: Section\,~\ref{Data} briefly introduces observation information about the stars and their spectra. In Section\,\ref{methods}, we show the details of the methods. We present the results and discussions in Section\,\ref{discussion}, and a summary is followed in Section\,\ref{summary}
\section{DATA} \label{Data}
\subsection{Planet-host stars}

\begin{figure*}
\centering
\includegraphics[width=\linewidth]{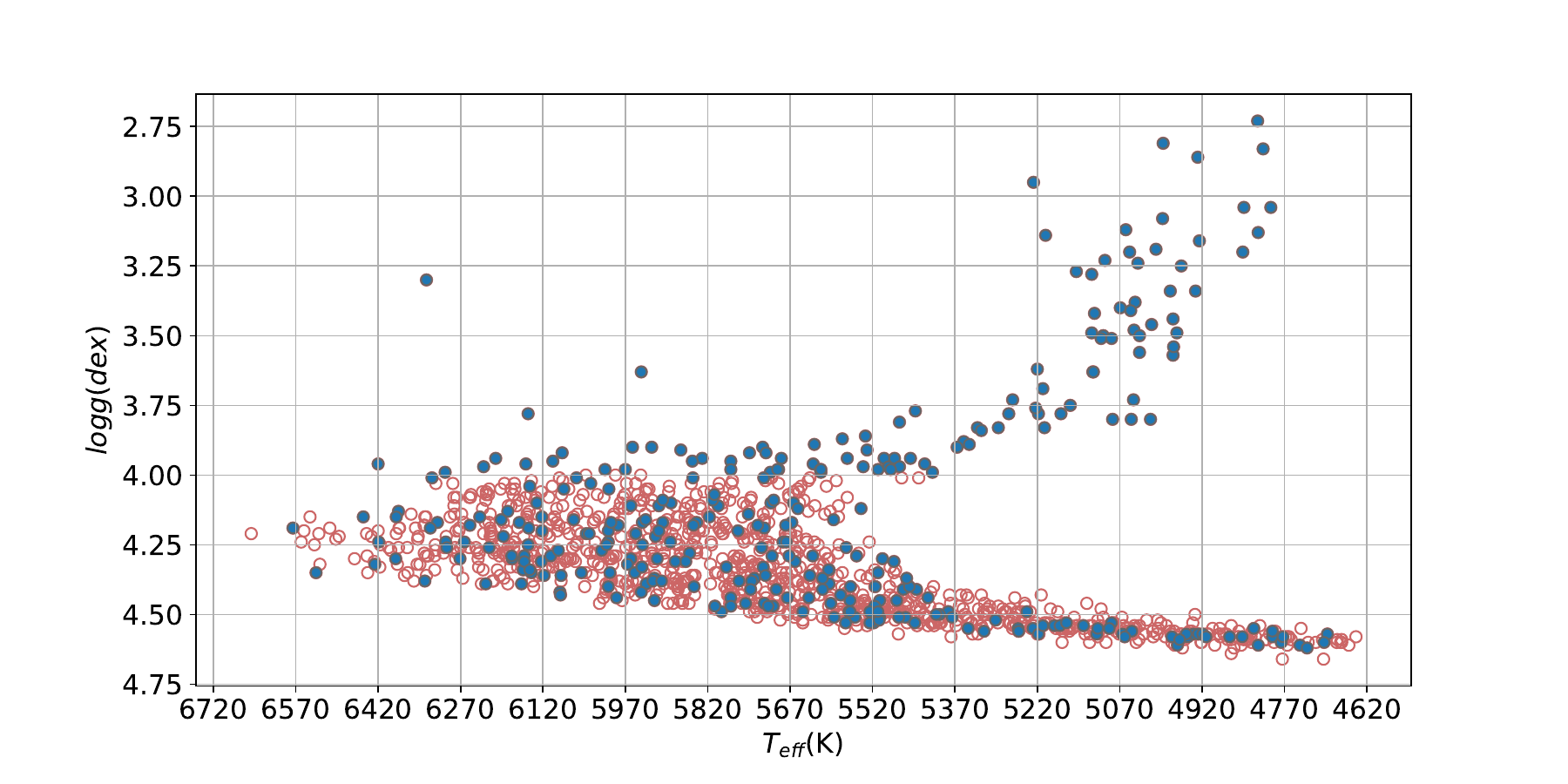}
\caption{The positions of the $1305$ HS on the H-R diagram are shown in the figure. Brown open circles represent the $1305$ plant-host stars from CKS and blue spots represent the $327$ adopted HS sample in this work.}
\label{fig1}
\end{figure*}

CKS published spectra of $1305$ KOIs and their atmospheric parameters \citep{2017AJ....154..108J}.
The CKS spectra were observed by the Keck/HIRES \citep{1994Sci...264..538W} with a resolution $\sim 50,000$ between $2012$ and $2015$. 
Our HS sample was selected from CKS. 
To make our HS sample fully covers the stellar parameter space of the CKS data, we first divided the CKS parameter space into a grid of $T_{\rm{eff}}$, $\log{g}$, and the range of the grid and separation of stellar parameters are as follows: $4,620 \le T_{\rm{eff}}\le 6,720$ with a grid interval of $150$\,K; $2.50 \le \log{g} \le 4.75$ with a grid interval of $0.25$\,dex. The grid of CKS parameter space can be seen in Fig.~\ref{fig1}. 

Then, we selected our HS sample from the CKS data. For the unevolved stars (we simply set $\log{g} = 4.0$ as the boundary of evolved and unevolved stars), we selected them by randomly picking $\sim 20\%$ stars out of each grid bin.
There are much less evolved stars than the unevolved ones in the CKS, and thus we include all the unevolved CKS stars in our HS sample. 
 
The above procedures result in $327$ HS sample (see Fig.~\ref{fig1}). We then cross-check these stars with the NASA exoplanet archive\footnote{https://exoplanetarchive.ipac.caltech.edu/} to ensure that all the selected HS are reliable. We found that $47$ stars are not found in the NASA exoplanet archive, and removed them from the sample. Finally, our HS sample consists of $279$ stars. 
The atmospheric parameters of our HS sample are directly adopted from the CKS result \citep{2017AJ....154..108J} except for the micro-turbulent velocity that was not provided by the CKS. The parameters' ranges of our HS sample are $4,690 \textless T_{\rm{eff}} \textless 6,600$\,K, $2.70 \textless \log{g} \textless 5.00$, $-0.55 \textless$ [Fe/H] $\textless 0.50$\,, respectively (details are shown in Table~\ref{tab1}). The mean signal-to-noise ratio (S/N) of our HS sample spectra is $\sim 50$.

\subsection{Isolated  Stars}
\label{sec:IS}

For comparison with the HS sample, we aim to select a ``pure" IS sample. Though it is impossible to define a $100\%$ isolated star at this stage, we take a step back to select stars with no detected variations of radial velocities (RV) after long-term RV monitoring.
We select the HS sample based on {\it Kepler} transit detections, while the IS sample is based on RV non-detections. Note that {\it Kepler} and RV methods have overlapping but not identical detection sensitivities (see, e.g., Figure 1 of \citealt{2021ARA&A..59..291Z} and related discussion in their review), and RV non-detections do not necessarily exclude the presence of {\it Kepler}-like planetary systems. Nevertheless, the frequency of {\it Kepler}-like planetary systems is about $30\%$ around FGK dwarfs \citep{2018ApJ...860..101Z}, so statistically speaking, the majority of our IS sample do not host {\it Kepler}-like systems.

In addition, for a coherent comparison, the IS sample is preferred to be observed with the same telescope and instrument under similar resolution.
Based on these two principles, we selected IS sample from $602$ stars used by \citet{2016ApJS..225...32B} and \citet{2020ApJ...896...64L}. \citet{2016ApJS..225...32B} produce $1617$ F, G, and K stars uniformly determined stellar properties by using the Keck/HIRES spectra. These stars are part of one or more radial velocity planet-search programs under the California Planet Survey. \citet{2020ApJ...896...64L} selected $602$ stars from \citet{2016ApJS..225...32B}, and verify reliability of stellar parameters by compare with other works \citep{2014A&A...562A..92D,2015A&A...574A..50J,2017A&A...606A..94D}.
We cross-matched these $602$ stars with the NASA exoplanet archive for selecting our IS sample.

To ensure that the stars in the comparison IS sample have similar stellar parameters with the stars in the HS sample, we employed a selection criterion based on the minimization of the difference in parameters. The difference, denoted as $\Delta$, is calculated using the following equation,
\begin{equation}
{\Delta = (\Delta T_{\rm{eff}}/1000)^2 + (\Delta \log\,g)^2 + (\Delta \rm [Fe/H])^2,}
\end{equation}
where $\Delta T_{\rm{eff}}$, $\Delta \log\,g$, and $\Delta \rm $[Fe/H] represent the difference of HS and IS on effective temperature, surface gravity, and metallicity, respectively.

\begin{figure*}
\centering
\includegraphics[width=\linewidth]{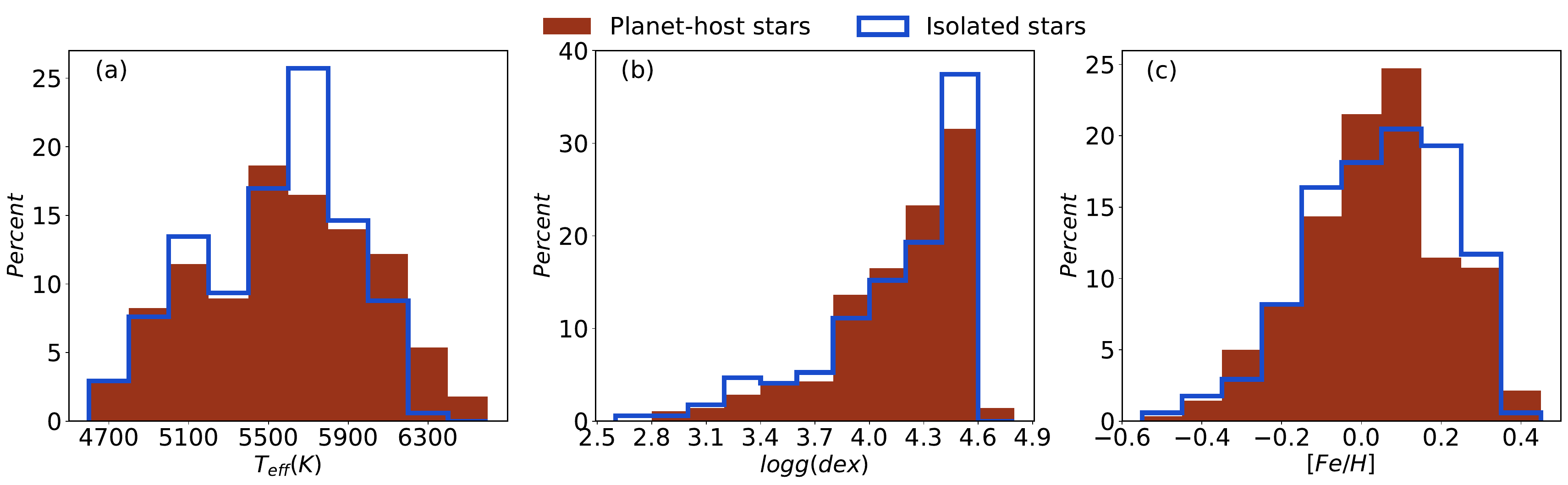}
\caption{Distribution of stellar sample. (a) Histogram showing $T_{\rm{eff}}$ for the $279$ HS and $171$ IS. (b) $\log{g}$ (c) [Fe/H]. }
\label{fig2}
\end{figure*}

By iterating over all the IS samples, we identified the star with the minimum $\Delta$ value relative to each star in the HS sample. Then the star with the minimum $\Delta$ value in $602$ IS is designated as the comparison IS for the corresponding HS.
This step gave us $171$ stars, and we present the comparison of the stellar parameter space between IS and HS samples in Fig.~\ref{fig2}.
Finally, our IS sample consists of $171$ IS, and the ranges of atmospheric parameters of the IS sample are as same as the HS sample.
The parameters' ranges of IS sample are $4,700 \textless T_{\rm{eff}} \textless 6,300$\,K, $2.5 \textless \log{g} \textless 5.0$, $-0.55 \textless$ [Fe/H] $\textless 0.50$, respectively (details are shown in Table~\ref{tab2}).

\section{Methods} \label{methods}
\subsection{Stellar Parameters} 

We directly adopted $T_{\rm{eff}}$, $\log{g}$ and [Fe/H] provided by \cite{2017AJ....154..108J} for our HS sample, and by \cite{2020ApJ...896...64L} for our IS sample, respectively.  
We noticed that \cite{2017AJ....154..108J} did not provide the micro-turbulence velocity ($\xi_{\rm t}$) which is necessary for our spectral synthesis method. 
Given that all our stars have metallicity [Fe/H]$\textgreater-1.0$, we calculated the $\xi_{\rm t}$ based on the empirical relations from literature for both HS and IS samples, which can be expressed as follows:
\\$T_{\rm{eff}} \le 
5000$\,K, $3.50\,\le\,\log{g}\,\le\,4.20$  \citep{2018MNRAS.481.2666G},
\begin{equation}
\xi_{\rm t}=1;
\label{equ2}
\end{equation}
$5000$\,K $\le T_{\rm{eff}} \le 5500$\,K, $3.50\,\le \log{g}\,\le\,4.20$ \citep{2021MNRAS.506..150B},
\begin{equation}
\xi_{\rm t}=1.1+1.6\times10^{-4}(T_{\rm{eff}}-5500\,K);
\label{equ3}
\end{equation}
$T_{\rm{eff}} \ge 5500$\,K or $\log{g}\ge4.20$ \citep{2018MNRAS.481.2666G},
\begin{equation}
\xi_{\rm t}=1.1+10^{-4}(T_{\rm{eff}}-5500\,K)+4\times10^{-7}(T_{\rm{eff}}-5500\,K)^2;
\label{equ4}
\end{equation}
$\log{g}\,\le\,3.50$ \citep{2018AJ....156..125H},
\begin{equation}
\xi_{\rm t}=10^{(0.226-0.0228\log{g}+0.0297(\log{g})^2-0.0113(\log{g})^3)};
\label{equ5}
\end{equation}
where empirical Eq.\,\ref{equ5} has no restriction on the $T_{\rm{eff}}$.

\subsection{Li abundance}

In this work, the Li abundance we used is expressed as A(Li), with the expression as: A(Li)=$\log(N_{\rm Li}/N_{\rm H})+12$, where $N_{\rm Li}$ and $N_{\rm H}$ represent the number densities of the lithium and hydrogen, respectively.

\begin{figure*}[t!]
\centering
\includegraphics[width=\linewidth]{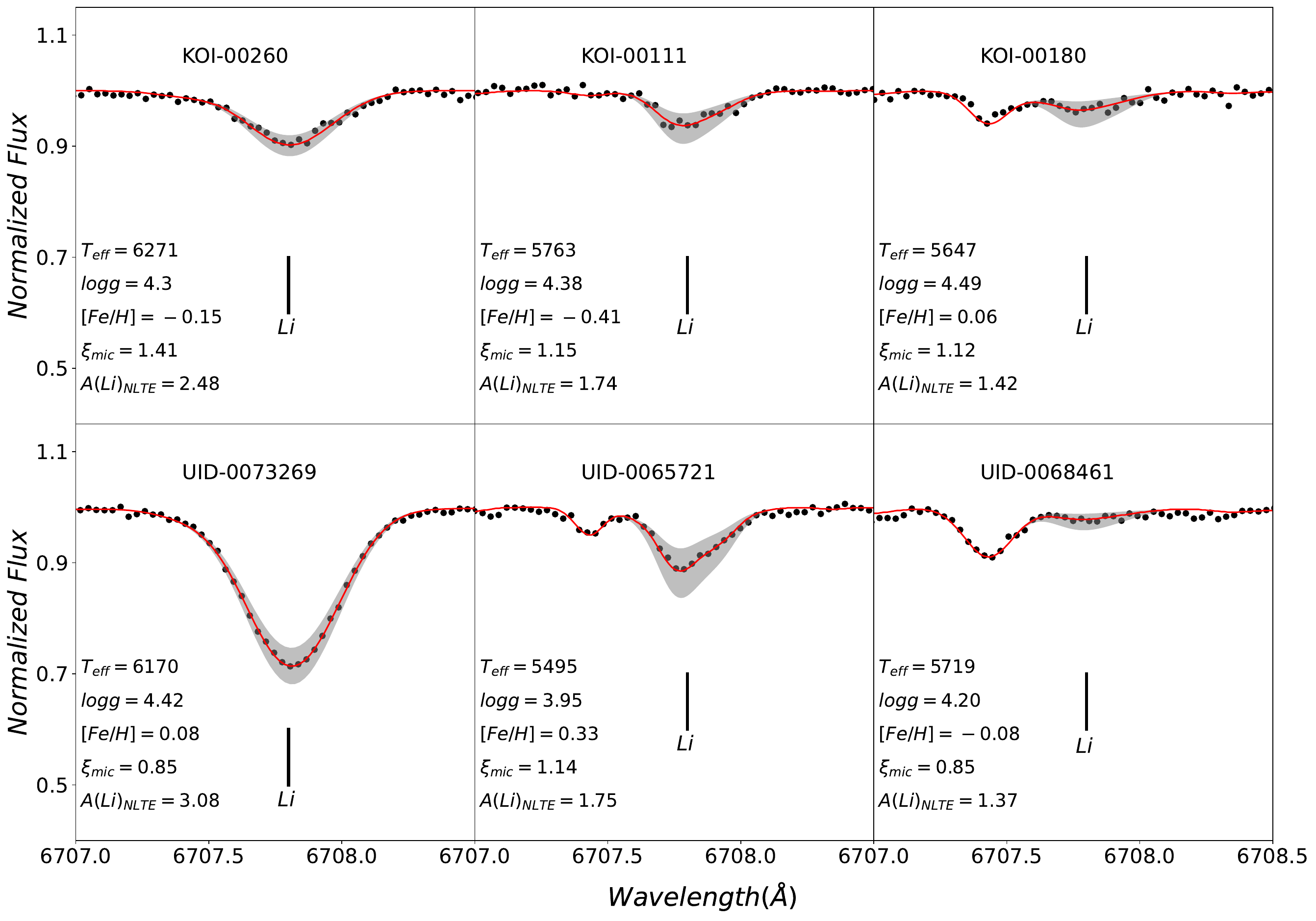}
\caption{Examples of spectral synthesis used in this work. The black dots are the observed spectra and the red lines are the theoretical spectra. From right to left, the gray areas represent synthesis spectra with $\rm{A(Li)_{{NLTE}}}\pm0.1\,dex$, $\rm{A(Li)_{{NLTE}}}\pm0.2\,dex$ and $\rm{A(Li)_{{NLTE}}}\pm0.3\,dex$, respectively.}
\label{fig3}
\end{figure*}

The abundance of Li was measured using the resonance \ion{Li}{1} lines at $6707.76$ and $6707.91$\,\AA. 
The measured Li abundance in this work was obtained by spectral synthesis method using the Spectrum Investigation Utility (SIU) code \citep{1999cezh.conf...66R}. We use the MARCS atmospheric models for abundance analysis \citep{2008A&A...486..951G}.
Both LTE and non-LTE abundance analysis were used in this study, and the LTE and non-LTE Li abundances are given in Tables~\ref{tab1} and \ref{tab2}. 
The atomic line data are adopted from \citet{2018NatAs...2..790Y}, see Tables~\ref{tab3}, and the atomic model of Li were adopted from \cite{2007A&A...465..587S}.
The van der Waals damping constant logC$_6$ for the \ion{Li}{1} is calculated from the \cite{1998MNRAS.296.1057B}. 
The Li abundances were derived from line-profile fitting.
In Fig.~\ref{fig3}, we show some fitting results for different Li abundances.

\begin{table*}
    \centering
    \tablenum{1}
    \caption{Li Abundance and Stellar Atmospheric Parameters of Planet-host stars}
    \begin{tabular}{llcccrcccc}
    \hline\hline
        {R.A}.& {Dec.} &  {{\it Kepler}-name} & {$T_{\rm{eff}}$(K)} & {$\log{g}$(dex)} &[Fe/H]& {$\xi_{\rm t}(km\,s^{-1})$} & {A(Li)$_{\rm{LTE}}$} & {A(Li)$_{\rm{NLTE}}$} \\ \hline
       292.24728 & 47.969521 & KOI-00002 & 6447 & 4.15 & 0.19 & 1.55 & 3.87 & 3.69 & ~ \\ 
        296.83725 & 48.239944 & KOI-00017 & 5667 & 4.17 & 0.34 & 1.13 & 0.94 & 0.94 & ~ \\ 
        283.25549 & 48.355232 & KOI-00046 & 5664 & 4.10 & 0.37 & 1.13 & 2.60 & 2.52 & ~ \\ 
        296.51059 & 42.54744 & KOI-00064 & 5354 & 3.88 & 0.09 & 1.09 & 0.99 & 1.02 & ~ \\ 
        287.698 & 42.338718 & KOI-00070 & 5507 & 4.45 & 0.10 & 1.10 & 0.90 & 0.92 & ~ \\ 
        285.67938 & 50.241299 & KOI-00072 & 5599 & 4.34 & $-$0.11 & 1.11 & 0.96 & 0.98 & ~ \\ 
        291.49722 & 42.728481 & KOI-00075 & 5922 & 3.90 & $-$0.04 & 1.21 & 0.96 & 0.96 & ~ \\ 
        290.42081 & 37.851799 & KOI-00084 & 5516 & 4.47 & $-$0.03 & 1.10 & 0.52 & 0.54 & ~ \\ 
        283.37482 & 43.788219 & KOI-00092 & 5917 & 4.37 & 0.08 & 1.21 & 2.31 & 2.27 & ~ \\ 
        295.43433 & 44.53112 & KOI-00099 & 4946 & 4.58 & $-$0.27 & 1.01 & 0.13 & 0.18 & ~ \\
        ... &  ... &  ... & ... & ... & ... &  ... & ...& ...  \\\hline
        
    \end{tabular}
    \label{tab1}
\end{table*}

\begin{table*}
    \centering
    \tablenum{2}
    \caption{Li Abundance and Stellar Atmospheric Parameters of Isolated stars}
    \begin{tabular}{llcccrccc}
    \hline\hline
        {R.A}.& {Dec.} &  {Name} & {$T_{\rm{eff}}$(K)} & {$\log{g}$(dex)} &[Fe/H]& {$\xi_{\rm t}(km\,s^{-1})$} & {A(Li)$_{\rm{LTE}}$} & {A(Li)$_{\rm{NLTE}}$} \\ \hline
       4.27313 & $\,-$1.65301 & HD1293 & 5114 & 3.19 & $-$0.37 & 0.85 & 0.53 & 0.58  \\ 
       4.67445 & $\,-$8.053 & HD1461 & 5739 & 4.34 & 0.16 & 1.15 & 1.24 & 1.24  \\ 
        7.72949 &$\quad$77.0194 & HD2589 & 5062 & 3.65 & $-$0.04 & 0.85 & 0.11 & 0.11 \\ 
        9.19506 & $\,-$24.50093 & HD3404 & 5339 & 3.81 & 0.20 & 1.09 & 1.85 & 1.86  \\ 
        10.29947 & \quad9.35506 & HD3861 & 6219 & 4.29 & 0.15 & 0.85 & 2.80 & 2.73  \\ 
        11.11104 & $\,-$26.51568 & HD4208 & 5639 & 4.50 & $-$0.29 & 1.12 & 0.61 & 0.64  \\ 
        11.36953 & $\,-$12.88081 & HD4307 & 5795 & 4.05 & $-$0.18 & 1.16 & 2.40 & 2.35  \\ 
        12.3615 & $\,-$23.21246 & HD4747 & 5305 & 4.56 & $-$0.24 & 1.10 & 0.56 & 0.62  \\ 
        15.27305 & \quad8.76937 & HD5946 & 5643 & 4.11 & 0.33 & 1.12 & 0.98 & 1.00 \\ 
        17.05202 & \quad21.97701 & HD6715 & 5624 & 4.48 & $-$0.21 & 1.12 & 0.69 & 0.70 \\ 
...&  ... &  ... & ... & ... & ... &  ... &...& ...\\\hline
        
    \end{tabular}
    \label{tab2}
\end{table*}

\begin{table*}
    \centering
    \tablenum{3}
    \caption{Li atomic data}
    \begin{tabular}{cccc}
    \hline\hline
        {Wavelength}& {Transition} &  {$\log{gf}$} & {$\log C_{6}$}  \\ \hline
       $6707.76$ &  $1s^22s-1s^22p$ & $-0.002$ & $-31.255$ \\
        $6707.91$ &  $1s^22s-1s^22p$ & $-0.299$ & $-31.255$ \\\hline
    \end{tabular}
    \label{tab3}
\end{table*}

The fitting quality is checked by eyes. 
For stars with high Li abundance and high S/N, we give the best-fit values.
While for the stars with low Li abundances, the profile of \ion{Li}{1} lines become hard to fit, and this situation is getting worse for spectra with low S/N.
For such stars, we only give the upper limit of the Li abundance.

\section{RESULTS and DISCUSSION}\label{discussion}

\subsection{Error Estimation and \\ Comparison with Other Results}

To estimate the errors of the derived Li abundances caused by the uncertainties of stellar parameters, we tested seven stars of different stellar parameters from our HS sample by varying the stellar parameters within their typical uncertainties.
We adopted the uncertainties of stellar parameters estimated by \cite{2017AJ....154..108J}, which are $\Delta T_{\rm{eff}} = \pm100$\,K, $\Delta \log{g} = \pm0.1$\,dex, and $\Delta$[Fe/H] $= \pm0.1$\,dex, respectively.
The $\xi_{\rm t}$ was calculated from the empirical relations, we estimated its uncertainty is $\Delta \xi_{\rm t} = \pm 0.1$\,km\,s$^{-1}$.
The results are shown in Table~\ref{tab4}. It needs to be pointed out that the Li abundance is more sensitive to the $T_{\rm{eff}}$ and [Fe/H].

\begin{table*}[h]
\tablenum{4}
\centering
     \caption{The Li Abundance Errors due to the Uncertainties in the Stellar Parameters} 
     \label{tab4}
\begin{tabular}{llccccccccc}
\hline \hline 
 ~ &\multicolumn{2}{c}{$T_{\rm{eff}}$} &\multicolumn{2}{c} {$\log{g}$}&\multicolumn{2}{c}{[Fe/H]} &\multicolumn{2}{c} {$\xi_{\rm t}$ }\\ \hline
 KOI-02828 &\multicolumn{2}{c}{4698K}&\multicolumn{2}{c}{4.6}&\multicolumn{2}{c}{-0.03}&\multicolumn{2}{c}{1km\,s$^{-1}$}\\
 Input error& $-$100K & $+$100K & $-$0.1 & $+$0.1 & $-$0.2 &$+$0.2 & $-$0.1km\,s$^{-1}$ & $+$0.1km\,s$^{-1}$ \\
$\Delta\,A(Li)$ & $-$0.14  &$+$0.13 & $+$0.01 & $-$0.01 & $+$0.09 & $-$0.1 & $-$0.01&  $+$0.01 \\ \hline
 KOI-00674 &\multicolumn{2}{c}{4973K}&\multicolumn{2}{c}{3.57}&\multicolumn{2}{c}{0.23}&\multicolumn{2}{c}{1km\,s$^{-1}$}\\
Input error& $-$100K & $+$100K & $-$0.1 & $+$0.1 & $-$0.2 &$+$0.2 & $-$0.1km\,s$^{-1}$ & $+$0.1km\,s$^{-1}$ \\
$\Delta\,A(Li)$ & $-$0.14  &$+$0.13 & $+$0.01 & $-$0.00 & $+$0.09 & $-$0.10 & $-$0.01&  $+$0.00 \\ \hline
 KOI-1413 &\multicolumn{2}{c}{5265K}&\multicolumn{2}{c}{3.73}&\multicolumn{2}{c}{-0.05}&\multicolumn{2}{c}{1.1km\,s$^{-1}$}\\
Input error& $-$100K & $+$100K & $-$0.1 & $+$0.1 & $-$0.2 &$+$0.2 & $-$0.1km\,s$^{-1}$ & $+$0.1km\,s$^{-1}$ \\
$\Delta\,A(Li)$ & $-$0.11  &$+$0.1 & $+$0.02 & $-$0.00 & $+$0.10 & $-$0.10 & $-$0.00&  $+$0.01 \\ \hline
 KOI-05929 &\multicolumn{2}{c}{5510K}&\multicolumn{2}{c}{4.35}&\multicolumn{2}{c}{-0.06}&\multicolumn{2}{c}{1.1km\,s$^{-1}$}\\
Input error& $-$100K & $+$100K & $-$0.1 & $+$0.1 & $-$0.2 &$+$0.2 & $-$0.1km\,s$^{-1}$ & $+$0.1km\,s$^{-1}$ \\
$\Delta\,A(Li)$ & $-$0.10 &$+$0.10 & $+$0.00 & $-$0.01 & $+$0.1 & $-$0.1 & $-$0.00&  $+$0.01 \\ \hline
 KOI-01316 &\multicolumn{2}{c}{5847K}&\multicolumn{2}{c}{4.01}&\multicolumn{2}{c}{0.3}&\multicolumn{2}{c}{1.18km\,s$^{-1}$}\\
Input error& $-$100K & $+$100K & $-$0.1 & $+$0.1 & $-$0.2 &$+$0.2 & $-$0.1km\,s$^{-1}$ & $+$0.1km\,s$^{-1}$ \\
$\Delta\,A(Li)$ & $-$0.09  &$+$0.08 & $+$0.00 & $-$0.01& $+$0.10 & $-$0.09 & $-$0.00&  $+$0.00 \\ \hline
 KOI-06102 &\multicolumn{2}{c}{6332K}&\multicolumn{2}{c}{3.3}&\multicolumn{2}{c}{-0.11}&\multicolumn{2}{c}{1.17km\,s$^{-1}$}\\
Input error& $-$100K & $+$100K & $-$0.1 & $+$0.1 &$-$0.2 &$+$0.2 & $-$0.1km\,s$^{-1}$ & $+$0.1km\,s$^{-1}$ \\
$\Delta\,A(Li)$ & $-$0.09  &$+$0.06 & $+$0.01 & $-$0.01& $+$0.1 & $-$0.1 & $-$0.00&  $+$0.01 \\ \hline
 KOI-02555 &\multicolumn{2}{c}{6144K}&\multicolumn{2}{c}{4.34}&\multicolumn{2}{c}{-0.06}&\multicolumn{2}{c}{1.33km\,s$^{-1}$}\\
Input error& $-$100K & $+$100K & $-$0.1 & $+$0.1 & $-$0.2 &$+$0.2 & $-$0.1km\,s$^{-1}$ & $+$0.1km\,s$^{-1}$ \\
$\Delta\,A(Li)$ & $-$0.07  &$+$0.07 & $+$0.00 & $-$0.01& $+$0.12 & $-$0.1 & $-$0.01&  $+$0.00 \\ \hline

\end{tabular}
\end{table*}

\begin{figure}[t]
\centering
\includegraphics[width=\linewidth]{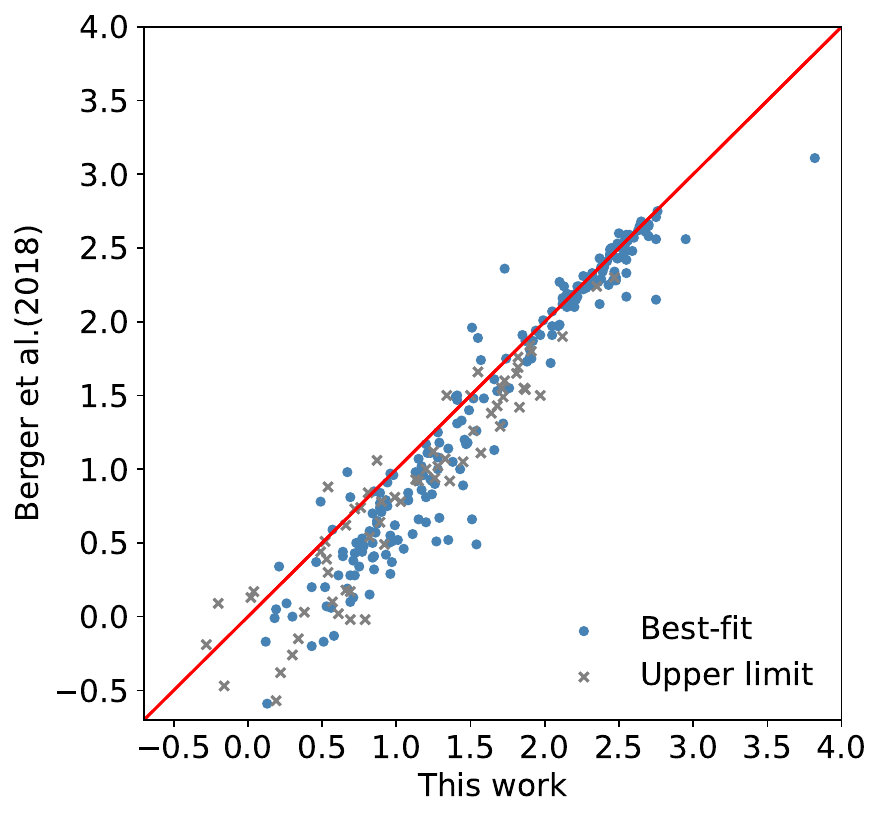}
\caption{ The Li abundance measured in this work is compared with those measured by \cite{2018ApJ...855..115B}. Blue circles represent best-fit Li abundance, and gray crosses represent the upper limit. The red line shows the $1$:$1$ relationship between the two parameters.}
\label{fig4}
\end{figure}

We also compared our results with literature.
For analyzing the distribution of Li abundance about planetary size, multiplicity, and orbital period, \cite{2018ApJ...855..115B} derived the Li abundances of $1305$ CKS stars by the equivalent width method.
We compare our results with \cite{2018ApJ...855..115B} in Fig.~\ref{fig4}, our A(Li)$_{\rm{LTE}}$ is generally agrees with \cite{2018ApJ...855..115B} in high Li abundance end.
This agreement shows that strong and well-defined spectral line profiles help to give more reliable abundances from both methods.
However, our A(Li)$_{\rm{LTE}}$ is larger than that of \cite{2018ApJ...855..115B} at the low Li abundance end.
Meanwhile, the scatter is also getting large at this range.
These differences, on one hand, are partly because our results consist of more upper-limit values at the low Li abundance end than that at the high Li abundance end.
On the other hand, they are possibly due to the continuum placement, as the abundance determination is more sensitive to the continuum at the low abundance end. 

\subsection{The non-LTE Effects}

\begin{figure}[t]
\centering
\includegraphics[width=\linewidth]{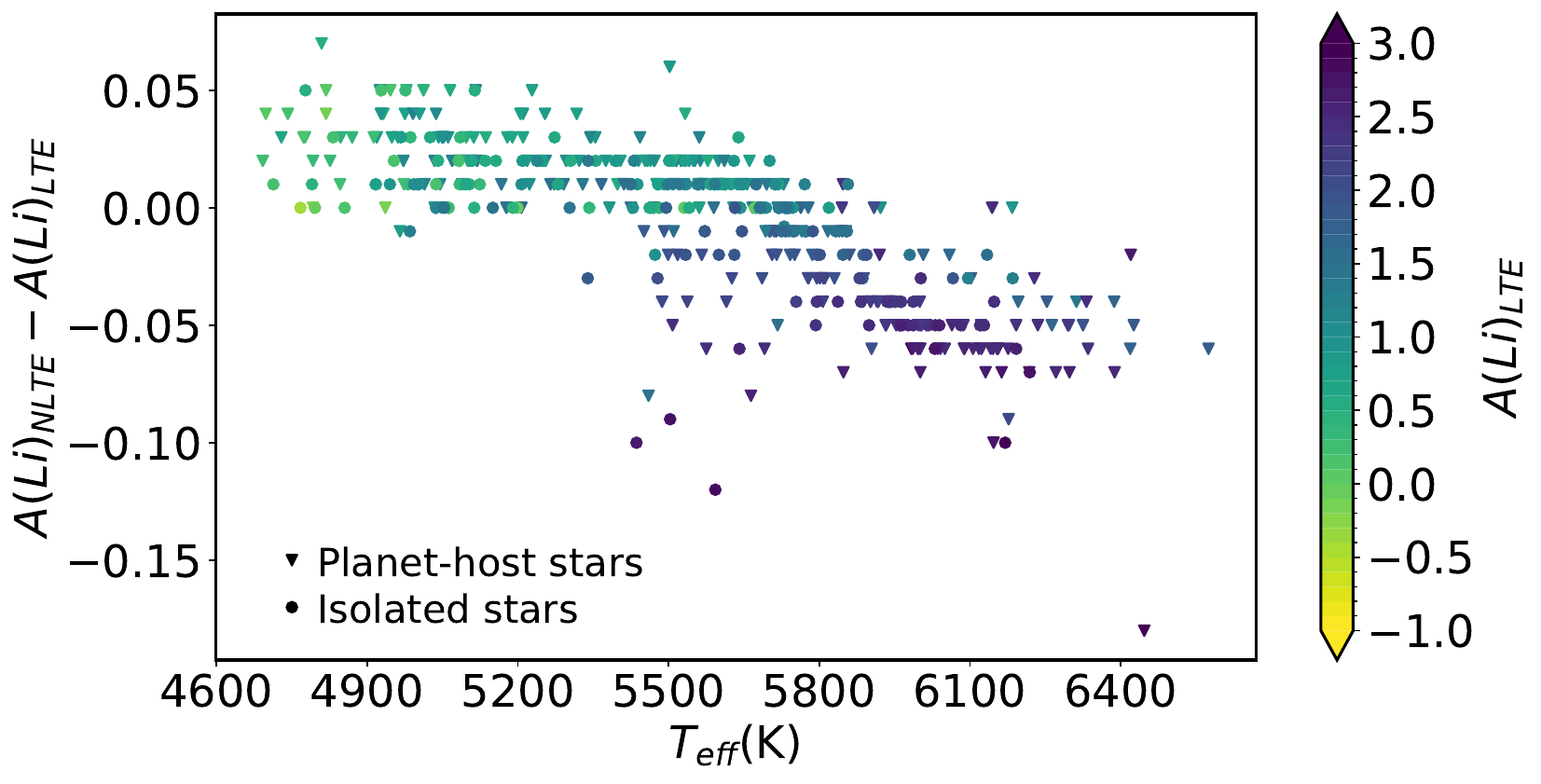} 
\caption{$\rm{A(Li)_{{NLTE}}}-\rm{A(Li)_{{LTE}}}$ of HS (triangle) and IS (points) versus $T_{\rm{eff}}$. The color-bar indicates $\rm{A(Li)_{{LTE}}}$}
\label{fig5}
\end{figure}

The non-LTE effects are sometimes important for the abundance determination \citep{2016ApJ...833..225Z}.
We present the non-LTE correction for Li abundance in Fig.~\ref{fig5}. The non-LTE correction is defined as A(Li)$_{\rm NLTE}-A(Li)_{\rm LTE}$. 
In general, the non-LTE correction of Li abundance is not large for the ``normal" stars. However, the non-LTE correction is more significant for stars with higher Li abundances, especially for objects with Li abundance higher than $2.5$\,dex.
The non-LTE correction is consistent with other non-LTE work such as \cite{2009A&A...503..541L} and \cite{2021MNRAS.500.2159W}.

\subsection{Differences of Li abundance in HS and IS}

\begin{figure*}
\centering
\includegraphics[width=150mm]{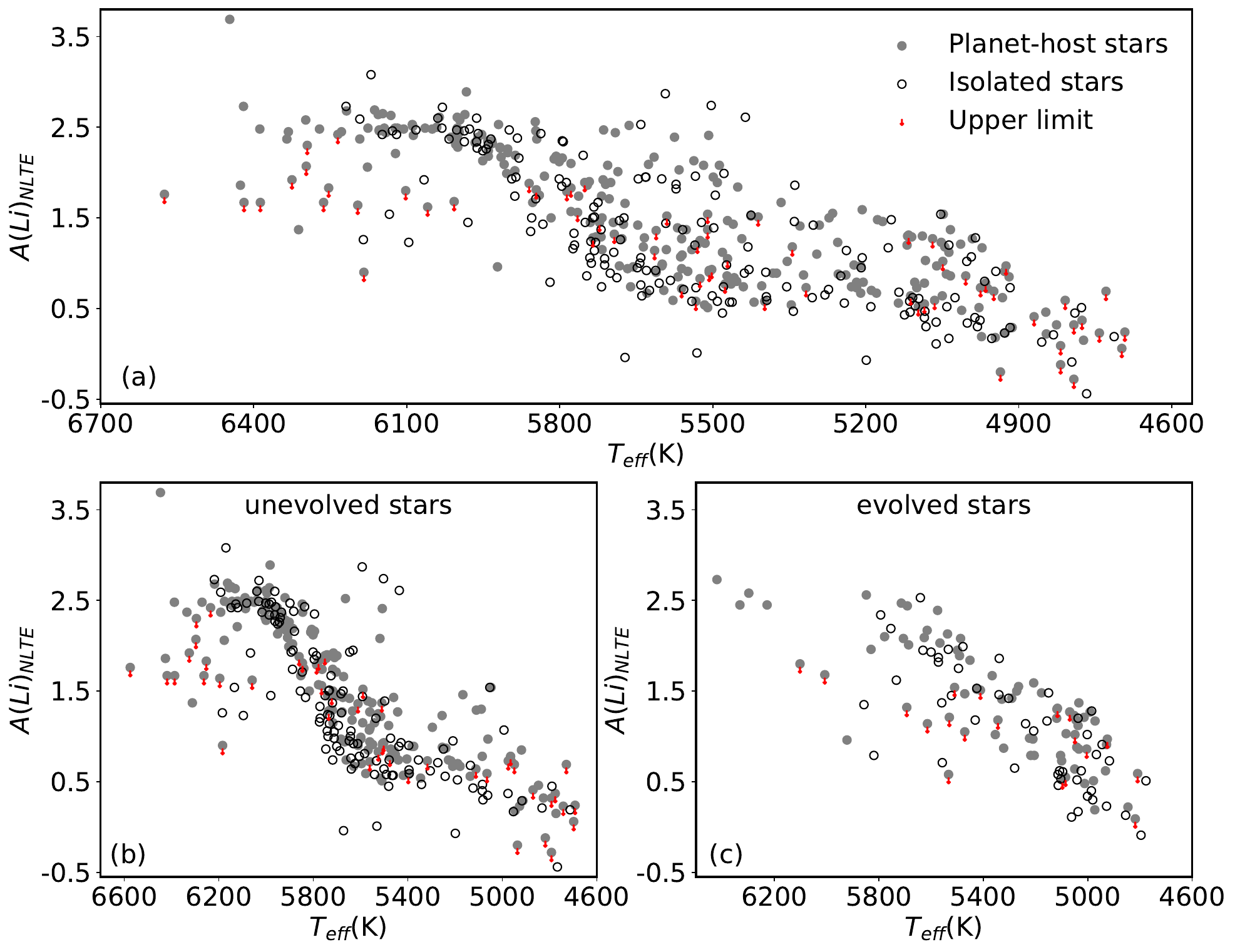}
\caption{$T_{\rm{eff}}$ vs. $A(Li)_{\rm {NLTE}}$. The figure shows the HS sample as gray filled dots and the IS sample as blank empty circles. The red down-arrows indicates stars with upper limit of Li abundance. Figure(a) shows the whole sample. The two panels (b) and (c) represent unevolved stars and evolved stars, respectively}
\label{fig6}
\end{figure*}

In this subsection, we discuss the differences in Li abundance in HS and IS samples. 
We show the Li abundances of HS and IS in Fig.~\ref{fig6}(a).
The overall distribution of Li abundances for HS and IS is consistent. Moreover, we do not see a significant deviation in the temperature range $5600$\,K $\le T_{\rm eff} \le 5900$\,K between HS and IS reported by Israelian et al. (2004).

\begin{figure}[hbt!]
\centering
\includegraphics[width=\linewidth]{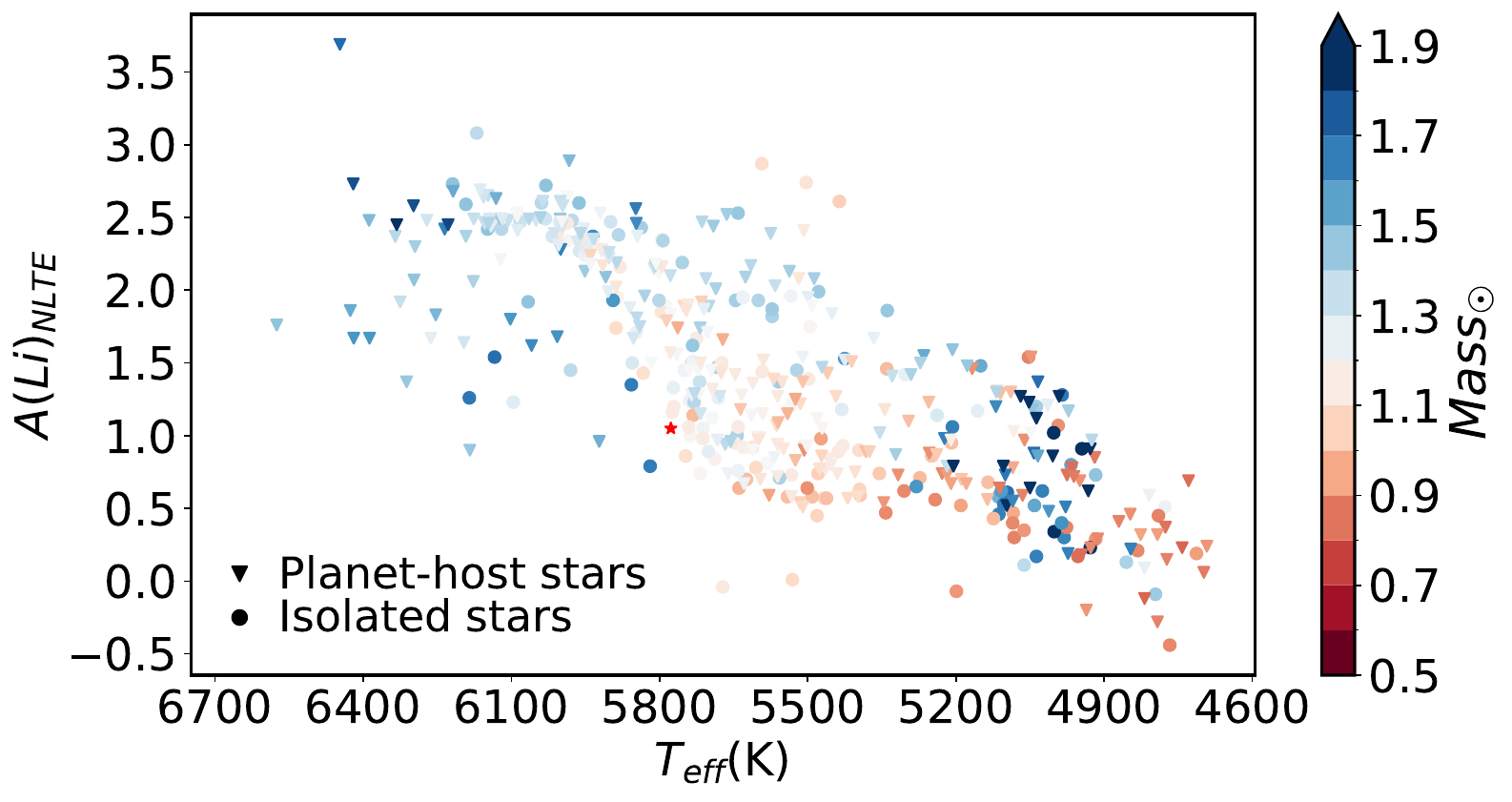}
\caption{$T_{\rm{eff}}$ vs. $A(Li)_{\rm{NLTE}}$; symbols have the same meanings as in Fig.~\ref{fig5}. The color-bar represents the mass of IS and HS.}
\label{fig7}
\end{figure}

Considering that stellar evolution can significantly alter the surface Li abundances, we divided our sample stars into evolved and unevolved stars using a simple criterion that whether $\log{g} \textgreater 4.0$\,dex. We compare the unevolved HS and IS samples in Fig.~\ref{fig5}(b) and evolved in Fig.~\ref{fig6}(c).
For the unevolved HS in the effective temperature range of $5500$ to $5800$\,K, there seems a slightly larger scatter on Li abundance than that of the IS with the same temperature range.
Given that the multiple mechanisms can affect the surface Li abundances of the unevolved stars, we can not identify that the larger scatter is due to the presence of exoplanets from the current data. 
Instead, we suspect that this scatter is due to the stellar mass.
In Fig.~\ref{fig7}, we show the stellar mass by colors for both HS and IS samples.
The stellar masses were obtained from the PAdova and TRieste Stellar Evolution Code (PARSEC) model \citep{1993A&AS..100..647B,2012MNRAS.427..127B}. We use the stellar parameters ($T_{\rm{eff}}$, $\log{g}$ and [Fe/H]) of $450$ stars to interpolate the PARSEC model, and determine their masses.
As seen in Fig.~\ref{fig7}, for the effective temperature range of $5500$ to $5800$\,K, stars with higher mass tend to have higher Li abundances.
This can be explained by the depth of the convection envelope.
Stars with higher mass in general have shallower convection envelopes, which prevent Li from being transported into deeper layers where temperatures are too high for Li to survive. Thus the depletion of Li in stars with higher masses is usually milder than that in stars with lower masses, leaving more Li on the surface.

For evolved stars, the strong depletion of the surface Li abundances is mostly due to the stellar evolution and the enhancement of mixing effects. This can also be seen in Fig.~\ref{fig6}, which shows the positions and Li abundances of the HS and IS on an H-R diagram. There is no sign that the presence of exoplanets strengthens such depletion effects from our data. On the contrary, exoplanets can be responsible for Li enrichment in the Li-rich evolved stars  \citep[e.g., see][and references therein]{2016ApJ...833L..24A,2019ApJ...880..125C,2021NatAs...5...86Y}. In summary, we did not find a significant difference between the Li abundance of HS and IS in a sample of $450$ stars with confirmed positive or negative detection of exoplanets, which are in agreement with previous studies \citep{2010A&A...519A..87B,2000MNRAS.316L..35R}.

\subsection{Effects of chromospheric activity}\label{sindex}


Since our IS sample was selected based on the non-detection of RV variations (details see Sec \ref{sec:IS}), such methods prefer stars with low chromospheric activity so that they have minimum effects on RV variations. In such a way, the IS sample may be biased towards lower chromospheric activity. In contrast, the stars in the HS sample were detected by the Kepler light curves, which does not strictly require low chromospheric activity for stars. Consequently, it is necessary to compare the chromospheric activity of two samples when considering the Li depletion.

Chromospheric activity can be traced by spectroscopic emission lines such as \ion{Ca}{2}\,H$\&$K and H$\alpha$ \citep{1981ApJS...45..635V,2015ApJ...809..157F,2024ApJ...976..243D}. Chromospheric activity can be quantified by canonical S-index \citep{1978ApJ...226..379W}, which is defined as:
 
\begin{equation}
S=8\alpha\frac{(H + K)}{(R + V)},
\label{equ6}
\end{equation} 
where H and K represent the integrated fluxes over a $1.09$\,$\rm{\AA}$ wide and centered on $3968.47$\,$\rm{\AA}$ and $3933.66$\,$\rm{\AA}$ for the \ion{Ca}{2}\,H$\&$K line cores, respectively. The other two parameters, R and V, are $20$\,$\rm{\AA}$ wide integrated fluxes and centered on $3901$\,$\rm{\AA}$ and $4001$\,\AA, respectively. We obtain the S-index of our samples by Eq.~\ref{equ6}. We set $\alpha$ = $2.4$ for these high-resolution spectra of HS and IS taken from Keck/HIRES following \citet{2018A&A...616A.108B}. It is noted that almost $90\%$ of our stars have very low chromospheric activities. The distributions of the S-index between HS and IS samples are similar to each other. Thus we conclude that the chromospheric activities of our IS and HS samples do not differ significantly.

\begin{figure*}
\centering
\includegraphics[width=\linewidth]{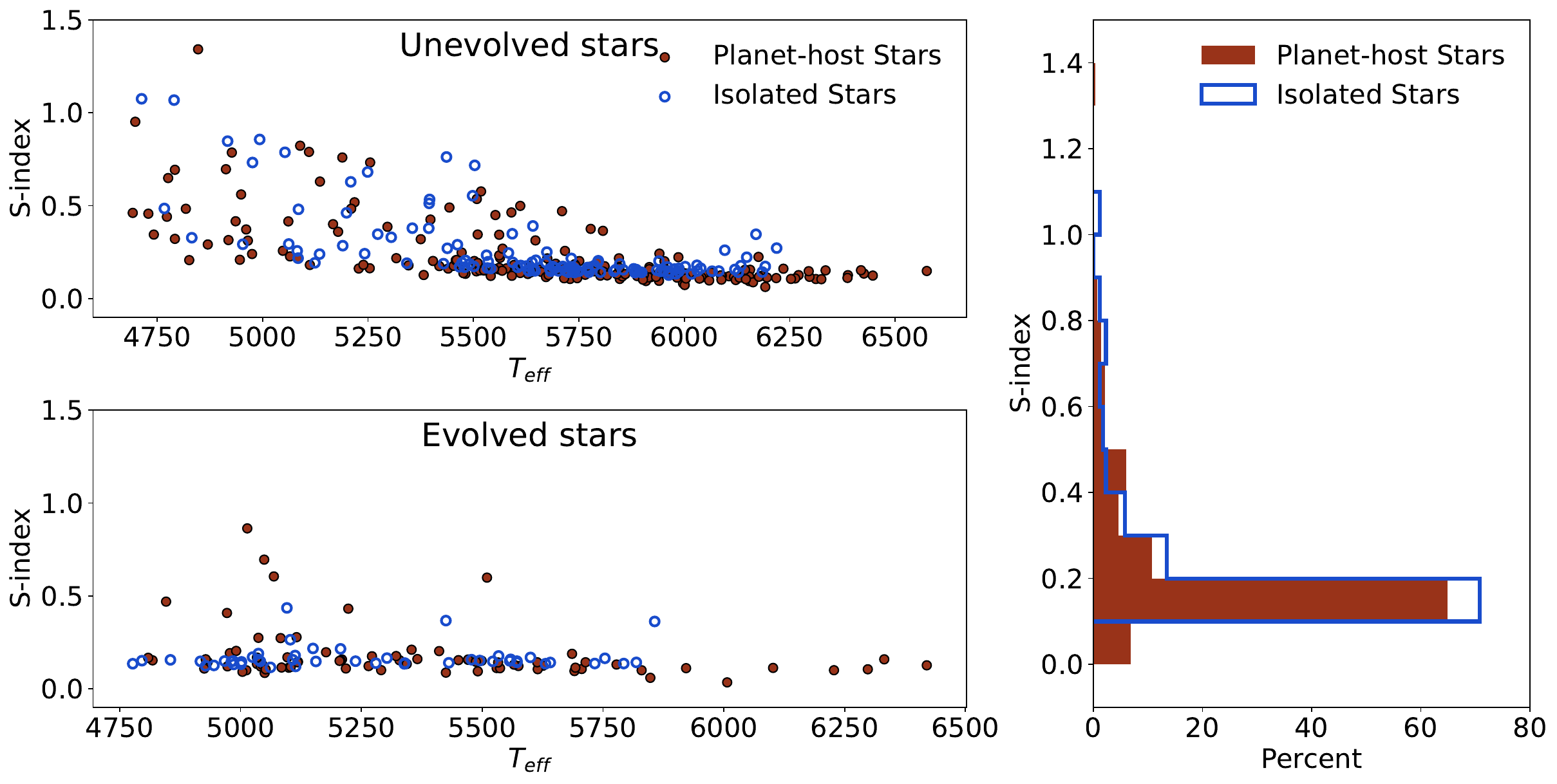}
\caption{Distribution of {S}-index. The left two panels represent unevolved stars and evolved stars, respectively. The HS sample is shown as brown filled circles and the IS sample is shown as blue empty circles. The histogram showing the frequency distribution of the S-index for HS(brown-filled bars) and IS(blue hollow bars) samples, with a bin width of $0.1$.}
\label{fig8}
\end{figure*}

\subsection{Discussion on stars in specific $T_{\rm{eff}}$ range}

Stars with $T_{\rm{eff}}$ ranging between ~$5600$\,K and ~$5900$\,K have been widely discussed in the literature. On one hand, these stars are partially considered as solar-analogs; on the other hand, they show significant Li depletion despite the depths of their convective zone is not sufficient to effectively dilute Li. Several studies suggest that stars hosting planets within this $T_{\rm{eff}}$ range exhibit more Li depletion compared to those without planets (e.g., I09). However, other works argue that this Li depletion cannot be distinguished between HS and IS samples \citep[e.g.,][L24 and references theirin]{2010A&A...519A..87B}. L24 studied a large sample consisting of 257 `Y stars' (stars with planet) and 1075 `N stars' (stars without planet), and find no evidence that the Li depletion differs between stars with our without planets.


\begin{figure*}[hbt!]
\centering
\includegraphics[width=\linewidth]{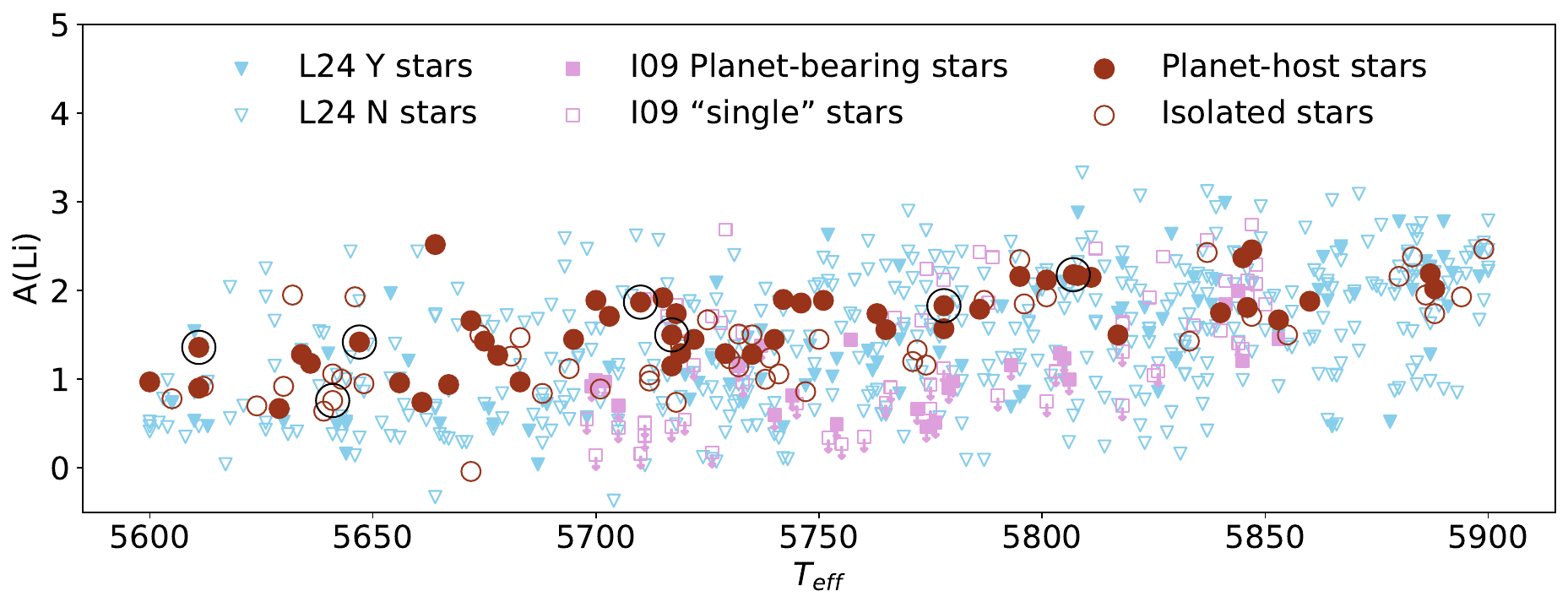}
\caption{A(Li) vs. $T_{\rm{eff}}$ for the stars from our sample (filled and empty dots), \citet[][filled and empty triangle]{2024A&A...684A..28L} and \citet[][filled and empty square]{2009Natur.462..189I}. Filled and empty markers represent stars with and without detected planets, respectively. The down-arrows indicates stars with upper limit of Li abundance and the black circles represent stars with higher S-index.}
\label{fig9}
\end{figure*}


To provide a more clear understanding of stars within this specific $T_{\rm{eff}}$ range in our sample, we present the distributions of Li abundances for this $T_{\rm{eff}}$ range, along with the Li abundances from I09 and L24 in Fig.~\ref{fig9}.  In general, it reveals a good agreement between our sample and L24's. In both L24 and our work, Li abundances of HS and IS (corresponding to L24's `Y stars' and `N stars', respectively) are hard to distinguish with each other, which is inconsistent with the result of I09. This is probably due to the small size of their `planet bearing stars' (corresponding to our HS) sample. Recent studies (e.g., L24 and this work) have expanded the Li abundance measurements of planet-host stars. As Fig.~\ref{fig9} shows, there is a significant number of planet-host stars exhibit Li abundances that is higher than those of I09. The new dataset indicates that the Li abundances of the HS and IS samples are still not distinguishable.





\begin{figure*}[hbt!]
\centering
\includegraphics[width=\linewidth]{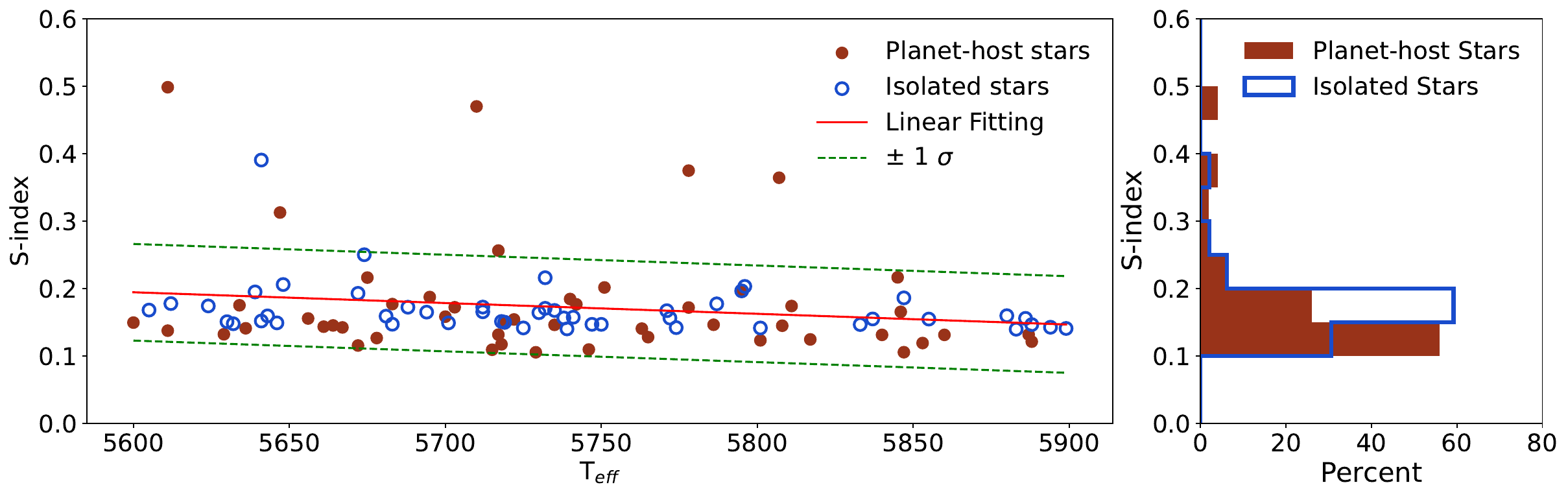}
\caption{Distribution of {S}-index in $5600$-$5900$\,K; symbols have the same meanings as in Fig.~\ref{fig8}. The green dashed lines represent 1$\sigma$ level of the fitted average.} 
\label{fig10}
\end{figure*}

To discuss whether that the presence of planet causes extra Li depletion in the specific $T_{\rm{eff}}$ range, the chromosperic activities must be considered. We show the chromosperic activities of the specific $T_{\rm{eff}}$ range in Fig.~\ref{fig10}. Most of the planet-host stars and isolated stars exhibit a low activity, which is similar with our main sample (see Fig.~\ref{fig8}). However, we noticed that there is a higher frequency of high activity stars in the HS sample of this $T_{\rm{eff}}$ range than that of IS sample. We defined stars whose S-index exceeds 1$\sigma$ level over the fitted average as high activity stars, and highlighted all seven (six HSs and one IS) of these stars with black circles in Fig.~\ref{fig9}. One can see those stars with low activities show similar Li abundances between HS and IS samples in the specific $T_{\rm{eff}}$ range. Stars with high activities tend to have higher Li abundances in Fig.~\ref{fig9}, which may be because these stars experienced a shorter time of Li depletion than those of stars with low activities. It is reasonable to assume that, over time, the Li abundances of these stars will decrease to an average level of low activity stars. Therefore, a higher frequency of high activity stars in the HS sample do not impact our conclusion.

\section{SUMMARY}\label{summary}

In this study, we analyzed a large sample of 450 stars with confirmed positive or negative detection of exoplanets, with temperatures ranging from $4600$ to $6700$\,K  and metallicity ranging from $-0.55$ to $0.50$\,dex.  The non-LTE effects were taken into consideration and it cannot be neglected for stars with A(Li) over $\sim 2.50$\,dex. We find that the overall distribution of Li abundances in both the HS and IS samples are generally consistent with each other.  For the unevolved stars, in the specific $T_{\rm{eff}}$ range($5600-5900$\,K), the Li abundances of the HS and IS samples are still not distinguishable. Although there is a higher frequency of high activity stars in the HS sample, this does not impact our conclusion.
For evolved stars, stellar evolution dominates the Li depletion. All these results suggest that there is no clear evidence that presence of planets could cause extra-Li depletion.

Given the complexity of Li depletion, to obtain a more reliable comparison of Li abundance for planet-host stars and isolated stars in the future, a much larger and more homogeneous sample is expected. For example, hundreds of planet-host stars and isolated stars can be selected from a volume-limited sample with better coverage of stellar parameters, especially in effective temperatures. 
In addition, with extensive stellar parameters and Li abundances of millions of stars \citep[e.g.,][]{2015MNRAS.449.2604D,2018MNRAS.478.4513B,2021MNRAS.506..150B,2022ApJS..260...45D,2024ApJS..271...58D,2024ApJS..273...18L}, the large-scale spectroscopic surveys also provide us opportunities to study such issues.

\begin{acknowledgments}
We thank the anonymous referee for the valuable comments and suggestions, which have significantly enhanced the clarity and depth of this paper. This study is supported by the National Key Basic R$\&$D Program of China No. 2024YFA1611903, the National Natural Science Foundation of China under grant No. 12173013, 12090040, 12090044, 12022304, 12373036, 11973052, the project of Hebei provincial department of science and technology under the grant number 226Z7604G, and the Hebei NSF (No. A2021205006). S.D. is supported by the National Natural Science Foundation of China (Grant No. 12133005). S.D. acknowledges the New Cornerstone Science Foundation through the XPLORER PRIZE. H.-L. Y. acknowledges support from the Youth Innovation Promotion Association of the Chinese Academy of Sciences. We acknowledge the China Manned Space Project for funding support of this study.  

\end{acknowledgments}
\bibliography{manuscript}{}
\bibliographystyle{aasjournal}

\end{document}